\documentstyle[12pt,epsfig]{article}

\def\L{\Lambda}

\def\d{\partial}

\def\/{\hfill\break}
\def\be{\begin{equation}}
\def\ee{\end{equation}}
\def\ba{\begin{eqnarray}}
\def\ea{\end{eqnarray}}
\def\half{{1\over2}}

\def\exv#1{\left\langle #1\right\rangle}
\def\mop#1{\mathop{\rm #1}\nolimits}
\def\Tr{\mop{Tr}}
\def\delv{\rlap{\lower-.5ex\hbox{--}}{\delta}}
\def\ket#1{\left| #1 \right\rangle}

\def\hphi{{\hat\Phi}}

\def\hh{{\hat H}}

\def\cO#1{{\cal O}\left(#1\right)}

\def\cL{{\cal L}}
\def\q{{\bf q}}
\def\p{{\bf p}}
\def\k{{\bf k}}
\def\x{{\bf x}}
\def\y{{\bf y}}
\def\z{{\bf z}}

\def\nn{\nonumber\\}
\def\fs{&&\hskip -0.6cm plus 0.1cm minus 0.1cm}

\def\DD{{\cal D}}

\def\adag{a^\dagger}
\def\om{\omega}
\def\e{\epsilon}

\date{} 
\title{\large\rm DESY 97-225 
  \hfill{\large\rm December 1997}\vspace*{2.5cm}\\ 
  {\LARGE\bf Classical Limit for Scalar Fields\\ 
   at High Temperature}} 
\author{W. Buchm\"uller and A. Jakov\'ac\\ 
  {\normalsize\it Deutsches Elektronen-Synchrotron DESY, 22603
    Hamburg, Germany}}

\begin{document}
\maketitle  

\begin{abstract}
\noindent
We study real-time correlation functions in scalar quantum field theories
at temperature $T=1/\beta$. We show that the behaviour of soft, long wave 
length modes is determined by classical statistical field theory. The loss 
of quantum coherence is due to interactions with the soft modes of the 
thermal bath. The soft modes are separated from the hard modes by 
an infrared cutoff $\L \ll 1/(\hbar\beta)$. Integrating out the hard modes 
yields an effective theory for the soft modes. The infrared cutoff $\L$
controlls corrections to the classical limit which are $\cO{\hbar\beta\L}$.
As an application, the plasmon damping rate is calculated. 
\end{abstract}

\setcounter{page}{0}
\thispagestyle{empty}
\newpage

\section{Introduction}

The properties of quantum field theories at high temperature are important
in order to understand the behaviour of matter under extreme conditions,
as they occur in astrophysics and cosmology. Static quantities at high
temperature characterize the thermodynamics of the system, in particular
the equation of state. Time-dependent quantities are relevant for the
study of non-equilibrium processes, such as the generation of a
cosmological baryon asymmetry.

Consider the partition function for a scalar field theory at temperature
$T=1/\beta$,
\be
Z\ =\ \int_{\beta} \DD\phi\ e^{-{1\over \hbar}\int_0^{\hbar\beta}
          d\tau d^3x \cL(\phi)}\quad,
\ee
where the functional integration is over periodic fields, $\phi(\x,\hbar\beta)
=\phi(\x,0)$. Formally, the limit $T\rightarrow \infty$ is closely related 
to the limit $\hbar \rightarrow 0$. This suggests that the high temperature 
behaviour of bosonic systems may be described by an effective classical field 
theory. For static quantities this reasoning leads to the dimensional reduction
program which has been successfully applied to the electroweak phase
transition in recent years \cite{ShapRub}. It has also been suggested that
the classical theory can be used to compute real-time correlation functions
at high temperature \cite{GriRub}. This is of particular interest for
sphaleron processes in the electroweak theory where, however, results
for the transition rate are still controversial \cite{ChSisim,ASY}.

Scalar field theories are much simpler than gauge theories and therefore
a useful starting point to study real-time processes at high temperature
\cite{paris}. This concerns the problem of ultraviolet divergencies
\cite{BoMcSm} as well as the question of quantum corrections to the classical 
limit \cite{Bod}. In classical $\phi^4$-theory one studies the classical 
correlation functions
\be\label{classcor}
\exv{\phi(\x_1,t_1)\ldots \phi(\x_2,t_2)}_{cl} =
{1\over Z} \int \DD\pi \DD\phi\ e^{-\beta H(\pi,\phi)}\ 
\phi(\x_1,t_1)\ldots \phi(\x_2,t_2)\;,
\ee
where
\be
H(\pi,\phi) = \int d^3x \left({1\over 2}\pi^2 + {1\over 2}(\nabla \phi)^2
 + {1\over 2} \mu^2\phi^2 + {1\over 4!} \lambda\phi^4\right)
\ee
is the classical Hamiltonian, and $\phi(\x,t)$ is the time-dependent
field which is obtained by integrating the classical equations of motion from 
initial conditions $\phi(\x,t_i)$, $\pi(\x,t_i)$ over which then a thermal 
average is performed. From the quantum thermal two-point function
one obtains the plasmon damping rate \cite{heinz}. As shown by Aarts and Smit
\cite{AaSm}, the damping rate is determined by the classical theory to leading
order in the coupling $\lambda$. This result is most easily obtained
from the retarded Green function, and it turns out that the whole imaginary
part of the self-energy is given by the classical theory in the 
high-temperature limit \cite{ourwork}. The classical limit for scalar field
theories has been further discussed in \cite{jak,weert}.

The quantity $\hbar\beta$ carries dimension. Hence, the question arises
which dimensionless quantity controls the classical limit. As is well
known, the spectral energy density of a relativistic Bose gas, i.e., Planck's
formula,
\be
de = {1\over \pi^2}{\hbar \om^3 \over e^{\hbar\beta\om}-1} d\om\; ,
\ee
turns into the classical Rayleigh-Jeans formula
\be
de = {1\over \pi^2} T \om^2 d\om
\ee
for small frequencies,
\be\label{classcon}
\hbar\beta\om \ll 1\quad .
\ee
Hence, the classical limit is obtained for low frequency, long wave length
modes. For high frequency modes there is no classical limit. 

We conclude that a consistent definition of the classical limit
requires the introduction of an infrared cutoff $\L \ll 1/(\hbar\beta)$ which 
separates `soft' modes with $\om < \L$ from `hard' modes with $\om > \L$.
The effective theory for the soft modes, where the hard modes have been
integrated out, can be approximated by a classical theory. For massive 
theories one has $m < \hbar\om < \hbar\L$. Since at finite temperature a 
plasmon mass $m^2 \sim \hbar\lambda T^2$ is generated, the condition 
(\ref{classcon}) implies $\hbar\lambda \ll 1$. Hence, a classical limit can 
only be obtained in the case of weak coupling.    

In the following sections we shall discuss the connection between the
finite-temperature quantum theory and the classical theory in detail. In
sect.~2 we construct an effective theory for the low frequency
modes. This is similar in spirit to the construction of an effective
average action \cite{wetter}, and in particular to the Wilson renormalization
group approach for thermal modes \cite{pietro}. Sects.~3 and 4 deal
with the infrared approximation which yields the classical limit up to
corrections $\cO{\hbar\beta\L}$ and quantum corrections. As an application, 
we discuss the evaluation of the plasmon damping rate in sect.~5, and sect.~6 
summarizes the main results. In the following we set $\hbar=1$.

\section{Low energy effective theory}

The basic quantities of finite-temperature field theory are the thermal 
averages of time-ordered products of field operators, 
\be\label{thgrf}
  \exv{{\rm T} \hphi(x_1)\ldots \hphi(x_n)}={1\over Z} \Tr\left(e^{-\beta\hh}
   \,{\rm T} \hphi(x_1)\ldots \hphi(x_n) \right) \; ,
\ee
which can be obtained by differentiation from the functional
\be
Z[j] = \Tr\left(e^{-\beta\hh}\,{\rm T}_c e^{i\int_c dx j \hphi} \right)\; .
\ee
Here $T_c$ denotes ordering along an appropriately chosen path $C$ in the
complex time plane \cite{LaWe}, and $\int_c dx j\hphi \equiv 
\int_c dt\int d^3x j(\x,t)\hphi(\x,t)$.

The dynamics of modes with energies $\omega$ below the temperature $T$ is
strongly affected by the thermal bath. In order to derive a low energy 
effective action for these modes one has to integrate out hard thermal 
modes. In the following we shall perform the part of the trace in 
Eq.~(\ref{thgrf}) which involves large momentum modes. This procedure is
similar to the Wilson renormalization group approach for thermal
modes \cite{pietro}.

A convenient basis for the evaluation of the trace is provided by the
coherent states of Bargmann and Fock (cf.~\cite{ItzZub}). They are 
constructed in terms of creation 
operators related to the field operator at some fixed time $t_i$,
\be
\ket{\eta} = N\ e^{\int_k \eta_\k a_\k^\dagger} \ket{0}\, ,
\ee
where $N$ is a normalization factor and
\be
 \int_\k = \int {d^3k\over (2\pi)^3 2\omega_\k}\; .
\ee
The coherent states are eigenstates of annihilation operators,
\be
a_\k \ket{\eta} = \eta_\k\ket{\eta}\, .
\ee
In the following we shall also use coherent states restricted to soft
modes,
\be
\ket{\eta_\L} = N_\L\ e^{\int_{|\k|<\L} \eta_\k a_\k^\dagger} \ket{0}\, ,
\ee
As discussed in the introduction, hard modes are defined by $\omega_\k > \L$,
where the separaton scale $\L$ has to be chosen in the range $m < \L
\ll \beta^{-1}$. 

We can now split the trace which gives the functional $Z[j]$
into soft and hard modes,
\be
Z[j] = \int\limits_{\omega<\L} \DD\eta\DD\eta^*\, Z_\L[j]\quad ,
\ee
where
\ba\label{zl}
Z_\L[j]&=&\int\limits_{\omega>\L} \DD\eta\DD\eta^*\,\exv{\eta\left|
    e^{-\half\beta\hh}\,e^{i\int_c dx j\hphi} e^{-\half\beta\hh}\right|
    \eta}\nn
&\equiv&{\rm Tr}_\L \left(
    e^{-\half\beta\hh}\,e^{i\int_c dx j\hphi} e^{-\half\beta\hh}\right)\; ,
\ea
and $\DD\eta\DD\eta^*\propto\prod_{\k} d\eta_{\k}d\eta^*_{\k}$. The dependence
of $Z_\L[j]$ on the soft modes $\eta_\k$, $\om_\k < \L$, will be discussed
in detail below.
Note, that we have written the trace in a symmetric form in order to maintain
the reality of expectation values of hermitian operators.
\begin{figure}[htbp]
  \begin{center}
    \leavevmode
    \epsfig{file=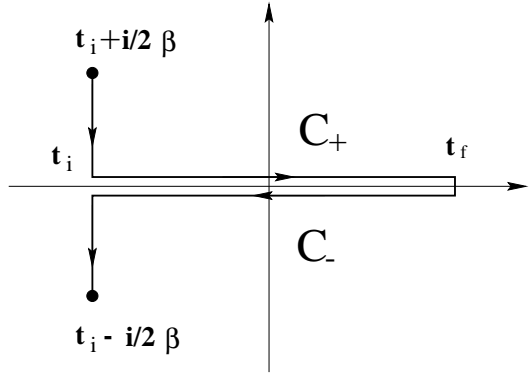,height=5truecm}    
  \end{center}
  \caption{\em Keldysh-type contour for real-time correlation functions}
  \label{fig:class_cont}
\end{figure}

The integral over the hard thermal modes yielding the functional $Z_\L[j]$
can be carried out in perturbation theory. Using the usual methods of real-time
perturbation theory \cite{LaWe} one easily derives the identity
\be\label{iden}
e^{-\half\beta\hh}\, {\rm T}_c e^{i\int_c dx j \hphi}\,e^{-\half\beta\hh} =
e^{i\int_c dx \cL_I\left({1\over i}{\delta\over \delta j}\right)}
e^{-\half\beta\hh_0}\, {\rm T}_c e^{i\int_c dx j \hphi_0}
\,e^{-\half\beta\hh_0}\, ,
\ee
where the subscript 0 refers to free fields which are related to interacting 
fields by a unitary transformation. $C$ is the Keldysh-type contour shown in 
Fig.~\ref{fig:class_cont},
\be
\label{mycontour}
  C:\, t_i+\frac i2\beta\to t_i \to t_f\to t_i\to t_i-\frac i2\beta.
\ee

From Eqs.~(\ref{zl}) and (\ref{iden}) one reads off,
\be\label{zl2}
Z_\L[j] = e^{i\int_c dx \cL_I\left({1\over i}{\delta\over \delta j}\right)}
Z_\L^{(0)}[j]\, ,
\ee
where
\be\label{zl0}
Z_\L^{(0)}[j] = \int\limits_{\omega>\L} \DD\eta\DD\eta^*\,\exv{\eta\left|
    e^{-\half\beta\hh_0}\,{\rm T}_c e^{i\int_c dx j\hphi_0} 
    e^{-\half\beta\hh_0}\right|
    \eta}.
\ee
In order to evaluate the integral over hard thermal modes we first express
the time-ordered product in terms of a normal-ordered product,
\be\label{phti}
{\rm T}_c e^{i\int_c dx j\hphi_0} =
e^{-\half\int_c dx_1 dx_2 j(x_1) G_c(x_1,x_2) j(x_2)}\,
:e^{i\int_c dx j\hphi_0}:\quad ,
\ee
where $G_c$ is the usual causal propagator \cite{NiSe},
\ba\label{propc}
G_c(x_1,x_2) &=& \int_\k G_c(\k,t_1,t_2)\ e^{i\k(\x_1-\x_2)} \nn
&=& \int_\k \left(\Theta(\tau_1-\tau_2)e^{-i\omega_\k (t_1 - t_2)}\right.\nn
&&\hspace{1cm}\left. + \Theta(\tau_2 -\tau_1)
e^{i\omega_\k (t_1-t_2)}\right)\ e^{i\k(\x_1-\x_2)}\; .
\ea
Here $\tau$, which parametrizes the contour $C$, is a real variable whereas
$t(\tau)$ is in general complex. 

For matrix elements of normal ordered products one easily proves the identity
\be 
  \exv{\eta_\k\left |e^{-\half\beta \hh_0} :f(\adag,a):
    e^{-\half\beta \hh_0} \right|\eta_\k} 
    = f(\eta_\k^* e^{-\half\beta\omega_\k},
              \eta_\k\, e^{-\half\beta\omega_\k})\, 
  \exv{\eta_\k\left|e^{-\beta \hh_0} \right|\eta_\k}.
\ee
This yields
\be\label{phno}
 \exv{\eta\left |e^{-\half\beta \hh_0} \ :e^{i\int_c j\hphi_0}:\
    e^{-\half\beta \hh_0} \right|\eta} = e^{i\int_c dx j\phi}
  \exv{\eta\left|e^{-\beta \hh_0} \right|\eta}\, ,
\ee
where the field $\phi$ is given by
\ba
\phi(\x,t) &=& \frac{\exv{\bar{\eta}\left|\hat{\phi}_0(\x,t)\right|\bar{\eta}}}
    {\exv{\bar{\eta}|\bar{\eta}}} \nn
 &=& \int_\k \left(\phi_\k \cos{\omega_\k (t-t_i)} + 
{\pi_\k\over \omega_\k}\sin{\omega_\k (t-t_i)}\right) e^{i\k\x}\; ,
\ea
with
\be
\phi_\k = e^{-\half\beta\omega_\k}\left( \eta_{-\k}^*+ \eta_\k \right)\quad , 
\quad\pi_\k = i \omega_\k e^{-\half\beta\omega_\k} \left( \eta_{-\k}^*- \eta_\k
\right)\; .
\ee
Here $|\bar{\eta}\rangle = \exp{(-{1\over 2}\beta \hat{H}_0)}|\eta\rangle$.
A straightforward calculation yields for the matrix element of the free 
Hamiltonian, 
\be\label{freeh}
 \exv{\eta\left |e^{-\beta \hh_0} \right|\eta} = 
\exp{\left(-{1\over 4}\int_\k {1\over n(\om_\k)}\left(
|\phi_\k|^2 + {1\over \omega_\k^2}|\pi_\k|^2\right)\right)}\, ,
\ee
where $n(\omega)$ is the Bose-Einstein distribution function
\be
n(\omega) = {1\over e^{\beta\omega}-1}\, .
\ee

The free functional $Z_\L^{(0)}[j]$ can now be obtained from Eqs.~(\ref{phti}),
(\ref{phno}) and (\ref{freeh}) by performing a Gaussian integration. The 
calculation gives the result
\be\label{zl02}
Z_\L^{(0)}[j] = e^{-\beta H_{0,\L}}\,
e^{-\half\int_c dx_1 dx_2 j(x_1) G_\L(x_1,x_2) j(x_2)}\,
e^{i\int_c dx j\phi_\L}\quad ,
\ee
where $H_{0,\L}$ and $\phi_\L$ are the free hamiltonian and the free field,
respectively, restricted to  soft modes,
\ba
e^{-\beta H_{0,\L}}&=&
\exv{\eta_\L\left|e^{-\beta \hh_0} \right|\eta_\L}\;, \\
\phi_\L(x)
&=& \frac{\exv{\bar{\eta}_\L\left|\hat{\phi}_0(\x,t)\right|\bar{\eta}_\L}}
    {\exv{\bar{\eta}_\L|\bar{\eta}_\L}} \nn
&=& \int_\k \Theta(\om - \L) \left(\phi_\k \cos{\omega_\k (t-t_i)} + 
{\pi_\k\over \omega_\k} \sin{\omega_\k (t-t_i)}\right) e^{i\k\x}\ ,
\label{fieldL}
\ea
with  $|\bar{\eta}_\L\rangle = 
\exp{(-{1\over 2}\beta \hat{H}_0)}|\eta_\L\rangle$.
Integration over the hard modes leads to a corresponding thermal 
contribution to the propagator,
\ba\label{propL}    
G_\L(x_1,x_2)&=&G_c(x_1,x_2) + G_{T,\L}(x_1-x_2)\quad, \\
G_{T,\L}(x_1-x_2)&=& 2 \int_\k n_\L(\omega_\k)
             \cos(\omega_\k (t_1-t_2)) e^{i\k(\x_1-\x_2)}\; ,
\ea
where $n_\L(\om) = \Theta(\om - \L) n(\om)$.
From Eqs.~(\ref{zl2}) and (\ref{zl02}) one obtains
\ba\label{zl3}
Z_\L[j] &=& e^{-\beta H_{0,\L}}\,
e^{i\int_c dx \cL_I\left({1\over i}{\delta\over \delta j}\right)}\,
e^{i\int_c dx j\phi_\L}\,
e^{-\half\int_c dx_1 dx_2 j(x_1) G_\L(x_1,x_2) j(x_2)} \nn
&=&e^{-\beta H_{0,\L}}\,e^{i\int_c dx j\phi_\L}\,\tilde{Z}_\L[j]\; ,
\ea
where
\be\label{ztild}
\tilde{Z}_\L[j] = 
e^{i\int_c dx \cL_I\left(\phi_\L+{1\over i}{\delta\over \delta j}\right)}\,
e^{-\half\int_c dx_1 dx_2 j(x_1) G_\L(x_1,x_2) j(x_2)}\, .
\ee
Note, that $Z_\L$ and $\tilde{Z}_\L$ are functionals of $j$ and $\phi_\L$, 
the soft modes, which play a role similar to a background field.

Using Eqs.~(\ref{zl}) and (\ref{zl3}) one can express the thermal Green
functions in a compact form. Changing variables from $\eta, \eta^*$ to
$\phi,\pi$ the result reads
\ba\label{npnpl}
  \exv{{\rm T} \hphi(x_1)\ldots \hphi(x_n)}&=&
  {1\over Z}{1\over i^n}\left.{\delta\over\delta j(x_1)}\ldots 
                           {\delta\over\delta j(x_n)}\, Z[j]\right|_{j=0}\nn
  &=&{1\over Z} \int\limits_{\omega<\L}\DD\phi\DD\pi\, e^{-\beta H_\L}
  \exv{{\rm T} \hphi(x_1) \ldots \hphi(x_n)}_\L
\ea
where
\be\label{hamL1}
H_\L = H_{0,\L}+H_{I,\L} = -{1\over \beta} \ln{Z_\L[0]}\; ,
\ee
and
\be\label{hamL2}
\exv{{\rm T} \hphi(x_1)\ldots \hphi(x_n)}_\L =
 {1\over Z_\L}{1\over i^n}\left.{\delta\over\delta j(x_1)}\ldots 
     {\delta\over\delta j(x_n)}\, Z_\L[j]\right|_{j=0}\, .
\ee

From these expressions one can easily recover the usual real-time perturbation
theory. The generating functional for the thermal Green functions is
obtained by integrating $Z_\L[j]$ over the soft modes. A Gaussian integration
yields,
\be\label{barg}
\int\limits_{\omega<\L} \DD\eta\DD\eta^*\, e^{-\beta H_{0,\L}}\,
e^{i\int_c dx j\phi_\L}\ \propto
e^{-\half\int_c dx_1 dx_2 j(x_1) \bar{G}_{T,\L}(x_1-x_2) j(x_2)}\; ,
\ee 
where
\be
\bar{G}_{T,\L}(x)= 2 \int_\k \bar{n}_\L(\omega_\k)
             \cos(\omega_\k t) e^{i\k\x}\; ,
\ee
where $\bar{n}_\L(\om) = n(\om) - n_\L(\om) = \Theta(\L -\om) n(\om)$.
Combining Eqs.~(\ref{zl3}) and (\ref{barg}) one obtains the familiar result
\be
Z[j] = e^{i\int_c dx \cL_I\left({1\over i}{\delta\over \delta j}\right)}\,
 e^{-\half\int_c dx_1 dx_2 j(x_1) G(x_1,x_2) j(x_2)}\; ,
\ee
where $G=G_\L + \bar{G}_{T,\L}$ is the usual thermal propagator \cite{NiSe}.
The perturbation series for Green functions with infrared cutoff $\L$ is
obtained from the perturbation series without cutoff by the following
procedure: one substitutes the propagator $G(x-y)$ by $G_\L(x-y)$, and one adds
contributions which are obtained by replacing a propagator $G(x-y)$ by
the product of background fields $\phi_\L(x)\phi_\L(y)$.

\section{Infrared approximation}

So far we have only reorganized the perturbative expansion, without making
any approximation. In the resulting low energy effective theory, however,
we have a small parameter, $\beta\Lambda\ll 1$, where $\L$ is the cutoff
which distinguishes between soft and hard modes. As discussed in the
introduction, in an expansion with respect to $\beta\Lambda$ the
leading term should correspond to the classical limit. We shall refer to
the leading term as the infrared (IR) approximation since, as we shall see, 
further approximations are necessary in order to obtain the classical limit. 

Consider the thermal Green functions with IR cutoff $\L$. From Eq.~(\ref{zl3})
one obtains
\be\label{dzl}
{1\over i}{\delta\over\delta j(x)} Z_\L[j] = e^{-\beta H_{0,\L}}\,
e^{i\int_c dx j\phi_\L}\,\left(\phi_\L(x) + 
{1\over i}{\delta\over \delta j(x)}\right)\,
\tilde{Z}_\L[j]\quad,
\ee
and therefore
\ba\label{npa}
\exv{{\rm T} \hphi(x_1)\ldots \hphi(x_n)}_\L &=& {1\over \tilde{Z}_\L}
\left(\phi_\L(x_1) + {1\over i}{\delta\over\delta j(x_1)}\right)\ldots\nn
&&\hspace{1.2cm}\left.\left(\phi_\L(x_n) + {1\over i}
{\delta\over\delta j(x_n)}
\right)\, \tilde{Z}_\L[j]\right|_{j=0}\, .
\ea
The right-hand side of the equation is the sum of products of $n-m$ fields
$\phi_\L$ ($0\leq m \leq n$) and $m$-point functions
\be\label{npb}
\tilde{\tau}(y_1,\ldots, y_m) =
{1\over \tilde{Z}_\L}{1\over i^n}\left.{\delta\over\delta j(y_1)}\ldots 
     {\delta\over\delta j(y_m)}\, \tilde{Z}_\L[j]\right|_{j=0}\, .
\ee

\begin{figure}[htbp]
  \begin{center}
    \leavevmode
    \epsfig{file=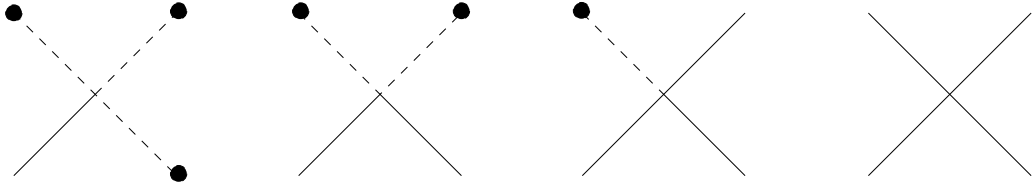,width=\hsize}
  \end{center}
  \caption{Vertices in the presence of the background field $\phi_{\L}$}
\label{fig:vertices}
\end{figure}

The $\tilde{\tau}$-functions can be computed as usual in perturbation
theory. According to Eq.~(\ref{zl3}), $\phi_\L(x)$ occurs as a background
field which leads to additional vertices (cf.~Fig.~\ref{fig:vertices}).The 
$\tilde{\tau}$-functions can be decomposed into disconnected
and connected parts (cf., e.g., \cite{zinn}),
\ba\label{npc}
\tilde{\tau}(y_1,\ldots, y_m)&=&\tilde{\tau}(y_1)\ldots\tilde{\tau}(y_m) \nn
&&+\tilde{\tau}(y_1,y_2)_c\, \tilde{\tau}(y_3)\ldots\tilde{\tau}(y_m)+
\mbox{permutations} \nn
&&+\ldots \nn
&&+\tilde{\tau}(y_1,\ldots, y_m)_c\quad.
\ea
For the tadpole one obviously has (cf.~(\ref{dzl})),
\be\label{1p}
\tilde{\tau}(y)=\exv{\hphi(y)}_\L - \phi_\L(y)\, .
\ee
From Eqs.~(\ref{npa})-(\ref{1p}) one obtains
\be\label{disc}
\exv{{\rm T} \hphi(x_1)\ldots \hphi(x_n)}_\L =
\exv{\hphi(x_1)}_\L \ldots \exv{\hphi(x_n)}_\L + \mbox{PC}\, ,
\ee
where PC denotes partially connected contributions which contain
connected $m$-point functions.

\begin{figure}[ht]
  \begin{center}
    \leavevmode
    \epsfig{file=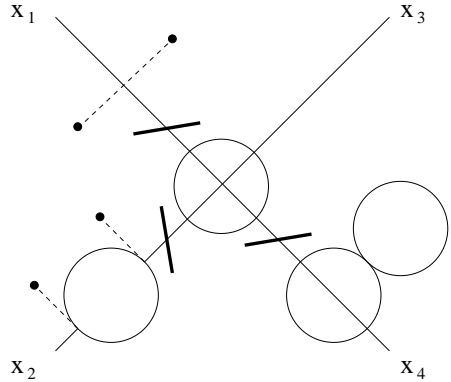,height=5truecm}
  \end{center}
  \caption{Contribution to $\tilde{\tau}(x_1,\ldots,x_4)_c$ with 
           cuts indicated}
\label{fig:compli}
\end{figure}

\begin{figure}[hb]
  \begin{center}
    \leavevmode
    \epsfig{file=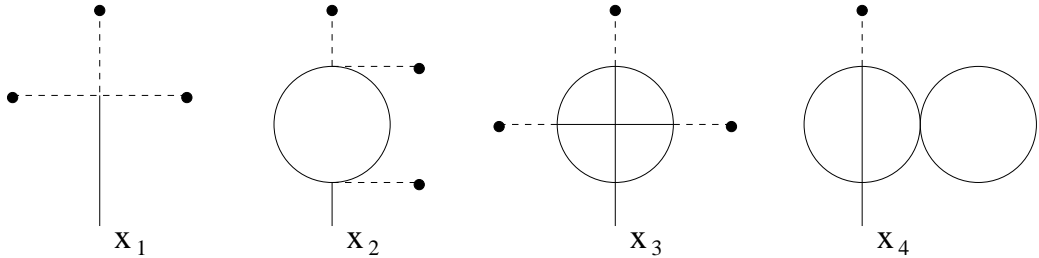,width=\hsize}
  \end{center}
  \caption{The tadpole diagrams corresponding to the cut components}
\label{fig:tadpole}
\end{figure}

Consider a particular contribution to PC with a connected $m$-point function 
$\tilde{\tau}(y_1,\ldots, y_m)_c$ (cf.~Fig.~\ref{fig:compli}). Substituting 
for $m-1$ lines
\be
G_\L(y-z) \rightarrow \phi_\L(y)\phi_\L(z)
\ee
yields a contribution to $\exv{\hphi(x_1)}_\L \ldots \exv{\hphi(x_n)}_\L$
(cf.~Fig.~\ref{fig:tadpole}). Hence, for each contribution to PC there exists a
corresponding term contained in the disconnected product
$\exv{\hphi(x_1)}_\L \ldots \exv{\hphi(x_n)}_\L$.

In order to obtain the complete thermal Green functions we still have to
integrate over the soft modes. As discussed in the previous section, the 
integration $\int_{\omega<\L}\DD\phi\DD\pi$ replaces products of background 
fields by the infrared propagator $\bar{G}_{T,\L}$ (cf.~Eq.~(\ref{barg})),
\be
\phi_\L(y)\phi_\L(z) \rightarrow \int\limits_{\omega <\L} 
\bar{G}_{T,\L}(\k,y_0-z_0)\ e^{i\k(\y-\z)}\, .
\ee 
For soft modes, with
\be
\omega < \L \ll T\; ,
\ee
the propagators $G_\L$ and $\bar{G}_{T,\L}$, which occur in the
disconnected and connected contributions, respectively, satisfy the
inequality,
\be
G_\L(\k,t) \sim {1\over \omega_\k} < \beta\L\  {T\over \omega_\k^2}
\sim \beta\L\ \bar{G}_{T,\L}(\k,t) \; .
\ee
This implies that the dominant contribution to thermal Green functions is
determined by the disconnected term $\exv{\hphi(x_1)}_\L \ldots 
\exv{\hphi(x_n)}_\L$ for soft external momenta, i.e., $|\x_i - \x_j| > 1/\L$.
From Eqs.~(\ref{npnpl}) and (\ref{disc}) one obtains in this case,
\be\label{class}
  \exv{{\rm T} \hphi(x_1)\ldots \hphi(x_n)}=
  {1\over Z} \int\limits_{\omega<\L}\DD\phi\DD\pi\, e^{-\beta H_\L}
  \exv{\hphi(x_1)}_\L \ldots \exv{\hphi(x_n)}_\L +  \cO{\beta\L}\;.
\ee
This equation is the main result of this paper. It is similar to the 
ansatz of classical statistical field theory, as described in the introduction.
There are, however, several differences with respect to the classical
theory. 

Consider first the hamiltonian $H_\L$, as defined in Eqs.~(\ref{hamL1}) and 
(\ref{hamL2}). The interaction part is, to leading order in the loop expansion,
\be
H_{I,\L} = - \int d^3x \cL_I(\phi(\x,t_i))\, + \, \cO{\beta\L} \; .
\ee
Higher order corrections can be calculated in perturbation theory and yield
further deviations from the classical Hamiltonian. Further, as we shall see 
in the next section, also the initial conditions for the fields 
$\exv{\hphi(\x,t)}_\L$ and $\exv{\hat{\Pi}(\x,t)}_\L=
\exv{\partial_t\hphi(\x,t)}_\L$ are different from the integration variables
in the functional integral (\ref{class}).

Hence, except for the momentum cutoff, the integration over $\phi$ and $\pi$
corresponds to an integral over all possible initial conditions. 
Eq.~(\ref{class}) differs from the naive classical limit by the presence of
the momentum cutoff $\L$, by quantum corrections to the classical hamiltonian
and by the deviation of the time evolution of $\exv{\hphi(x)}_\L$ from the
classical evolution.

It is very interesting that in the high-temperature limit the coherence of the 
quantum fields is lost to leading order in $\beta\L$. In Eq.~(\ref{class}) the
expectation
value of the product of field operators has been replaced by the corresponding
product of expectation values, and the time ordering has therefore lost its 
meaning. This is an effect of the thermal bath where the propagator of the 
soft modes is dominated by the thermal part $\bar{G}_{T,\L}$.

\section{Time evolution of expectation values}

\begin{figure}[htbp]
  \begin{center} 
  \leavevmode 
     \epsfig{file=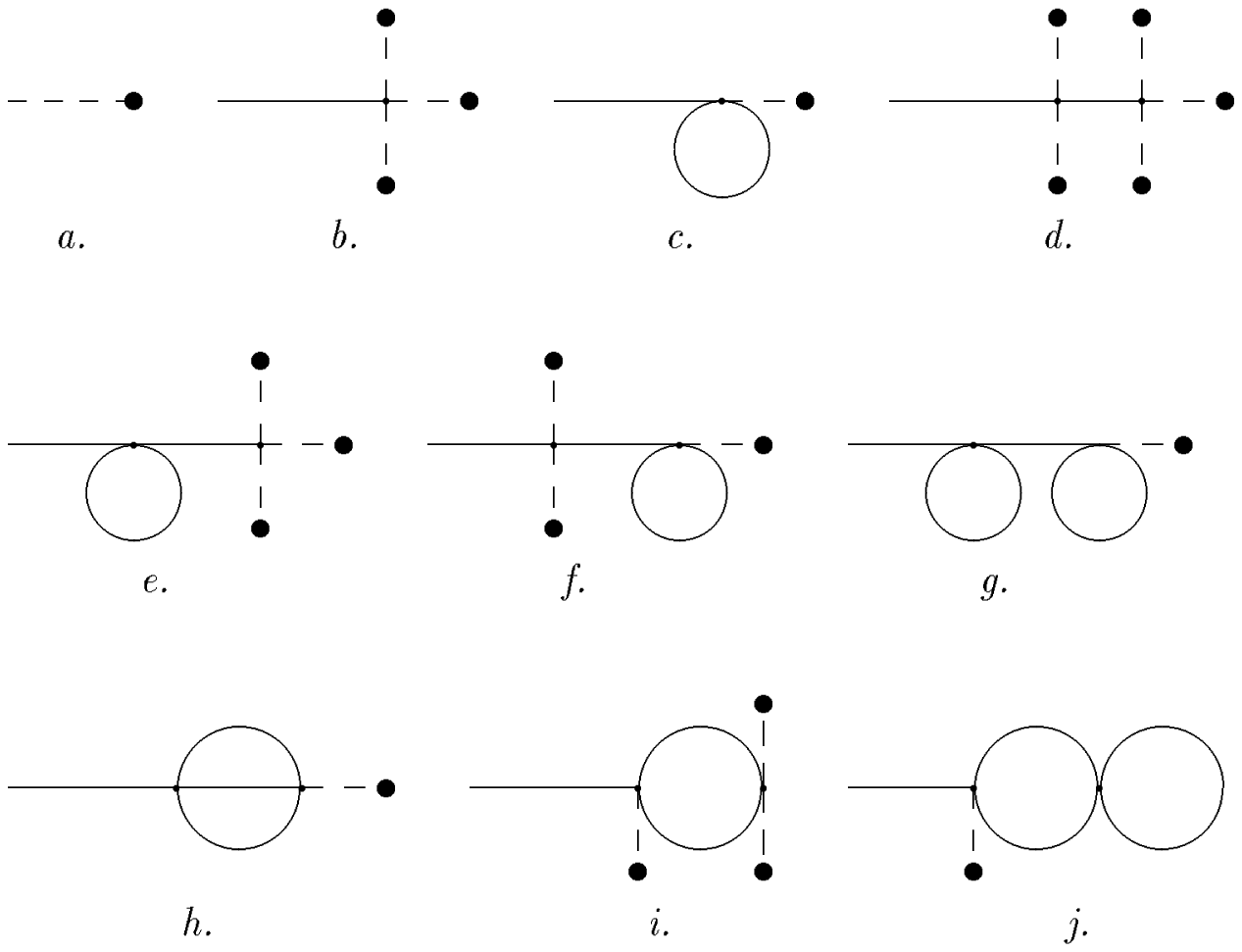} 
  \end{center}
  \caption{\em Contribution to $\exv{\hphi(x)}_\L$ up to two loops}
\label{fig:kernels}
\end{figure}

In order to understand the connection between the infrared approximation 
and the classical approximation one has to study the time evolution of
field expectation values in the low energy effective theory. 
The time dependent field
\be\label{exv}
  \exv{\hphi(x)}_\L = {1\over Z_\L} \Tr_\L\left(e^{-\beta\hh}\,\hphi(x)\right) 
\ee
can be evaluated in perturbation theory. From Eqs.~(\ref{zl3}) and 
(\ref{ztild}) one reads off,
\ba\label{pertfie}
\exv{\hphi(x)}_\L &=& \left(\phi_\L(x) + {1\over i}
{\delta\over \delta j(x)}\right)\nn
&&\hspace{0.5cm}\left.
e^{i\int_c dx \cL_I\left(\phi_\L+{1\over i}{\delta\over \delta j}\right)}\,
e^{-\half\int_c dx_1 dx_2 j(x_1) G_\L(x_1,x_2) j(x_2)}\right|_{j=0}\, .
\ea
To be specific, consider a quartic self-interaction,
\be
\cL_I(\phi) = {\lambda\over 4!}\ \phi^4\; .
\ee
The corresponding Feynman diagrams contributing to $\exv{\hphi(x)}_\L$ are 
shown in Fig.~\ref{fig:kernels} up to two loops.

The kinetic operator for a mode with momentum $\k$ is
\be
    K_0(\k)=\frac{d^2}{d^2t}+\omega_\k^2\; .
\ee
From the definitions (\ref{fieldL}) and (\ref{propL}) one reads off that the 
background field $\phi_\L$ and the propagator $G_\L$ satisfy the equations 
($t_i<t<t_f$),
\ba
  \fs K_0(\k)\phi_\L(\k,t)=0\; , \nn
  \fs K_0(\k)G_\L(\k,t)=-i\delta(t)\; .
\ea
Furthermore, one easily obtains for the field $\exv{\hphi(x)}_\L$, up to 
terms $\cO{\lambda^3}$ (cf.~Fig.~\ref{fig:kernels}), 
\be\label{queqm}
  K_0\,\exv{\hphi}_\L=-{\lambda\over 3!}\exv{\hphi}_\L^3- 
  (D_c+D_h+D_j)\exv{\hphi}_\L - \exv{\hphi}_\L\,D_i\,\exv{\hphi}_\L^2\; .
\ee
Here the $D_n$ are non-local kernels which give the contributions of the 
corresponding 1PI diagrams in Fig.~\ref{fig:kernels}.
Eq.~(\ref{queqm}) is similar to the classical equation of motion. The only
difference is that the coupling constants $m^2$ and $\lambda$ have
been replaced by {\em non-local} kernels. Note also, that higher orders
in the loop expansion will generate higher powers of $\exv{\hphi}_\L$
in Eq.~(\ref{queqm}).

Also important are the initial conditions which solutions of Eq.~(\ref{queqm})
have to satisfy. From Eq.~(\ref{pertfie}) it is clear that
\be
\exv{\hphi(\x,0)}_\L \neq \phi_\L(\x,0)\; .
\ee
Instead, one finds for the Fourier coefficients of both fields,
\ba\label{ini_cond}
\fs \exv{\hphi(\k,0)}_\L=\phi_\k + \cO{\lambda\beta\L}\; ,\nn
\fs \exv{\d_t\hphi(\k,t)}_\L\biggr|_{t=0}=\pi_\k + \cO{\lambda\beta\L}\;.
\ea
Hence, the integration variables in the infrared approximation (\ref{class})
agree with the initial conditions of the field $\exv{\hphi(x)}_\L$ only up
to corrections $\cO{\lambda\beta\L}$.

Consider now the quantum corrections to the classical field equation.
At one loop we have to consider the diagrams Figs.~\ref{fig:kernels}$c$
and \ref{fig:kernels}$i$. The first term is local and leads to an effective
thermal mass,
\be
m_{eff}^2=m_B^2+ \frac\lambda2\int_\k\ G_\L(\k,0)
= m_R^2+ {\lambda \over 24}\ T^2 \left(1 - {6\over \pi^2}\ \beta\Lambda
 \right)=m_\L(T)^2\; .
\ee
Here $m_B$ and $m_R$ are the bare mass and the renormalized mass of the
zero temperature 4D theory, respectively. The thermal mass $m_\L(T)$ is
well known from finite-temperature perturbation theory. 

Diagram \ref{fig:kernels}$i$ yields the contribution
\be\label{lamnl}
\exv{\hphi(x)}_\L\ D_i\exv{\hphi}_\L^2(x) = 
-i {\lambda^2\over 4}\ \exv{\hphi(x)}_\L\
 \int_c dx'\ G_\L(x,x')^2\ \exv{\hphi(x')}_\L^2\; ,
\ee
where the contour integral can be carried out by means of the two-component
formalism (cf.~appendix).
The divergent contribution is again renormalized as in the zero temperature
4D theory. The contributions \ref{fig:kernels}$b$ and \ref{fig:kernels}$i$
can be combined to an effective non-local coupling. A straightforward
calculation yields the result (cf.~appendix),
\be
\lambda_{eff}(x-x')=\lambda(T)\delta(x-x') \,+\,\lambda_{nonloc}(x-x'), 
\ee
where $\lambda(T)$ is the temperature dependent coupling familiar from
dimensional reduction \cite{JakPat},
\be
\lambda(T) = \lambda_R(\mu)\left(1 - {1\over 4\pi^2} \ln{\mu\over T}\right) 
              + \cO{\beta\L}\; ,
\ee
and
\newpage
\ba\label{nlocal}
\lambda_{nonloc}(\k,t-t')\fs=\frac{3\lambda^2}{4} \, \int\!\frac{d^3\q}
{(2\pi)^3} \,\frac1 {\omega_\q \omega_{\k-\q}}\nn 
\fs\Biggl[ \frac{ 1 +n_\L(\omega_\q) + n_\L(\omega_{\k-\q})}
{\omega_\q+\omega_{\k-\q}} \cos(\omega_\q+\omega_{\k-\q})(t-t')\,+\nn
\fs+\, \frac{ n_\L(\omega_{\k-\q}) -n_\L(\omega_\q)} {\omega_\q-
\omega_{\k-\q}}\, \cos(\omega_\q-\omega_{\k-\q}) (t-t') \Biggr]
\Theta(t-t'){\partial\over \partial t'}\nn
\fs\, +\,\cO{\beta\Lambda}\; .
\ea
To obtain this result we have performed a partial integration in $t'$
and discarded the contribution at $t'=-\infty$ (cf.~\cite{BoyVeg}). 
From the upper limit of integration $t'=t$ one obtains a contribution which is
local in space up to corrections $\cO{\beta\Lambda}$ \cite{JakPat}. The
important infrared contribution to the nonlocal part of the coupling was
previously found by Boyanovski et al. in a non-equilibrium context 
\cite{BoyVeg}.

\section{Retarded Green function}

An interesting illustration of the classical limit is the calculation of
the plasmon damping rate \cite{AaSm}, which is most easily evaluated by
considering the retarded Green function \cite{ourwork}. 

The retarded Green function of the low energy effective theory is given by
\be
i D_\L^R (x_1,x_2) = {1\over Z} \Tr_\L\left(e^{-\beta \hat{H}}
        \left[\hphi(x_1),\hphi(x_2)\right]\right)\Theta(t_1-t_2)\;.
\ee
It can be obtained from the time-ordered Green function by means of the
following relations,
\be\label{dt12}
i D_\L^R (x_1,x_2) = \left(D^>_\L(x_1,x_2) - D^<_\L(x_1,x_2)\right)
\Theta(t_1-t_2)\;,
\ee
and
\be
\exv{{\rm T}\hphi(x_1)\hphi(x_2)}_\L = 
D^>_\L(x_1,x_2)\ \Theta(t_1-t_2) + D^<_\L(x_1,x_2)\ \Theta(t_2-t_1)\; .
\ee
As discussed in Sect.~3, the time-ordered Green function is the sum of
a disconnected and a connected piece,
\be
\exv{{\rm T}\hphi(x_1)\hphi(x_2)}_\L = \exv{\hphi(x_1)}_\L \exv{\hphi(x_2)}_\L
  + D_\L^c(x_1,x_2)\; .
\ee
In the retarded Green function the disconnected piece drops out. 

\begin{figure}[htbp]
  \begin{center}
    \leavevmode
    \epsfig{file=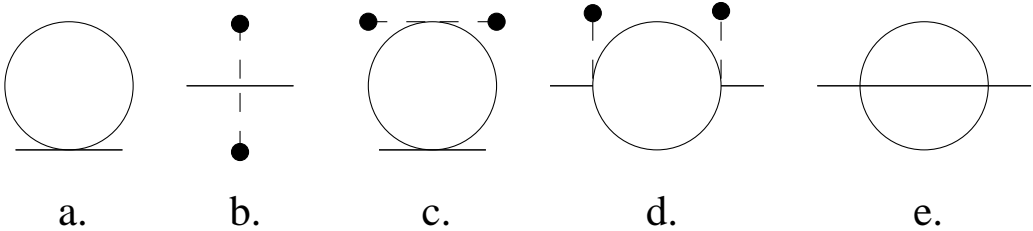,width=\hsize}
  \end{center}
  \caption{Contributions to the self-energy $\Pi$ up to ${\cal
    O}(\lambda^2)$.}
  \label{fig:selfen}
\end{figure}

As usual the connected piece satisfies a Dyson-Schwinger equation,
\be\label{dyson}
D_\L^c(x_1,x_2) = G_\L(x_1-x_2) + \int dy_1 dy_2\ G_\L(x_1-y_1) \Pi(y_1-y_2)
                                   D_\L^c(y_2,x_2)\; .
\ee
The contributions to the self-energy are shown in Fig.~\ref{fig:selfen} up
to order $\lambda^2$. To leading order the self-energy is local. The
contribution Fig.~\ref{fig:selfen}a is removed by a mass counter term and
Fig.~\ref{fig:selfen}b is the interaction with the background field,
\be
\Pi^{(1)}(y_1,y_2) = -i \lambda\ \phi_\L^2(y_1)\  \delta(y_1-y_2)\; .
\ee 

The free propagator $G_\L$ may also be split into two contributions which
describe propagation forward and backward in time, respectively,
\be\label{gt12}
G_\L(x_1-x_2) = G^{>}_\L(x_1-x_2)\ \Theta(t_1-t_2) +
                G^{<}_\L(x_1-x_2)\ \Theta(t_2-t_1) \; .
\ee
From Eq.~(\ref{propc}) and (\ref{propL}) one reads off,
\ba
G^>_\L(x_1-x_2) &=& \int_\k e^{-ik\cdot(x_1-x_2)} + G_{T,\L}(x_1-x_2) \nn
G^<_\L(x_1-x_2) &=& \int_\k e^{ik\cdot(x_1-x_2)} + G_{T,\L}(x_1-x_2) \; .
\ea
The corresponding retarded Green function reads
\ba
D^R_\L(x_1-x_2) &=& -i \left(G^{>}_\L(x_1-x_2 - G^{<}_\L(x_1-x_2)\right) 
                \Theta(t_1-t_2) \nn
              &=& - 2 \int_\k\ e^{i\k(\x_1-\x_2)}\sin(\omega_\k (t_1-t_2))
                   \Theta(t_1 - t_2)\; .
\ea
                
Inserting Eqs.~(\ref{dt12}) and (\ref{gt12}) in the Dyson-Schwinger
equation (\ref{dyson}) one obtains  two coupled integral equations for
$D^>_\L(x_1,x_2)$ and $D^<_\L(x_1,x_2)$. For the retarded Green function
$D_R$ one finds,
\be
D^R_\L(x_1-x_2) = G_R(x_1-x_2) + {1\over 2}\lambda \int dy\ G_R(x_1-y)
                \phi^2_\L(y) D^R_\L(y-x_2)\; .
\ee
This equation is identical with the equation for the classical retarded
Green function $H_R$ obtained in \cite{ourwork}. Instead of the classical
field $\phi(y)$ now the expectation value $\phi_\L(y)$ appears. 

\begin{figure}[htbp]
  \begin{center}
    \leavevmode
    \epsfig{file=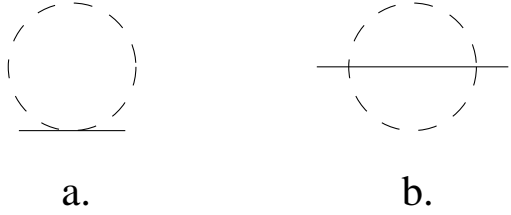,height=2.8truecm}
  \end{center}
  \caption{Contributions to the classical self-energy $\bar\Pi$ up to
    ${\cal O}(\lambda^2)$.}
  \label{fig:clasself}
\end{figure}

In order to obtain the complete retarded Green function one has to
perform the integration over the soft modes,
\be
D_R(x_1,x_2) = \int\limits_{\omega < \L}\DD\phi \DD\pi\ 
                 e^{-\beta H_\L}\ D^R_\L(x_1-x_2)\; .
\ee
This means, one has to evaluate the quantities
\be
\int\limits_{\omega < \L}\DD\phi \DD\pi\  e^{-\beta H_\L} 
\phi^2_\L(x_1) \ldots \phi^2_\L(x_{2n})\; .
\ee
This is completely analogous to the integration over initial conditions
in \cite{ourwork}. For two fields one has to leading order in $\lambda$ 
(cf. Eq.~(\ref{barg})),
\be
\int\limits_{\omega < \L}\DD\phi \DD\pi\  e^{-\beta H_\L} 
\phi_\L(x_1) \phi_\L(x_2)
\propto \bar{G}_{T,\L}(x_1-x_2) \; .
\ee
For the products of $2n$ fields one obtains,
\ba
\int\limits_{\omega < \L}\DD\phi \DD\pi\  e^{-\beta H_\L} 
\phi^2_\L(x_1) \ldots \phi^2_\L(x_{2n})\hspace{5cm} \nn
\quad = 2^n\ \bar{G}^2_{T,\L}(x_1-x_2)\ldots \bar{G}^2_{T,\L}(x_{2n-1}-x_{2n})
  + \mbox{permutations}\; .
\ea 
One easily verifies that the full retarded Green function again satisfies
a Dyson-Schwinger equation,
\be
D_R(x_1-x_2) = G_R(x_1-x_2) + \int dy_1 dy_2\ G_R(x_1-y_1)\
               \bar{\Pi}(y_1,y_2)\ D_R(y_2,x_2)\; .
\ee
The leading contributions to $\bar{\Pi}(y_1,y_2)$ are shown in 
Fig.~\ref{fig:clasself}. 
The first contribution is subtracted by a mass counter term. The second
contribution reads,
\be
\bar{\Pi}^{(2)}(y_1,y_2) = \lambda^2\ G_R(y_1-y_2)\ 
                           \bar{G}_{T,\L}^2(y_1-y_2)\; .
\ee
This yields
\ba\label{se}
\bar{\Pi}^{(2)}(\p,\om) &=& \int dt d^3x\ e^{i\om t} e^{-i\vec{p}\vec{x}}
                             \ \bar{\Pi}^{(2)}(t,\vec{x}) \nn
&=& {1\over 2} \lambda^2 T^2 \sum_{\eta_1,\eta_2,\eta_3} \eta_1
\int\limits_{\om_1,\om_2 < \L} d\Phi(\p) \nn
&& \hspace{2cm} {1\over \om_2\om_3}
   {1\over \om + \eta_1\om_1 + \eta_2\om_2 + \eta_3\om_3 + i\e}\, ,
\ea 
where $\eta_i = \pm 1$, $i=1,\ldots,3$ and
\be
d\Phi(\p) = {d^3q_1\over (2\pi)^3 2\om_1}
 {d^3q_2\over (2\pi)^3 2 \om_2} {d^3q_3\over (2\pi)^3 2 \om_3}
 (2\pi)^3 \delta({\bf p}-{\bf q_1}-{\bf q_2}-{\bf q_3}) \, .
\ee
From Eq.~(\ref{se}) one obtains for the imaginary part of the self-energy
($\om > 0$),
\ba\label{gamc}
\bar{\Gamma}^{(2)}(\p,\om)&=&{\pi\over 2}\lambda^2 T^2 
  \int\limits_{\om_1,\om_2 < \L} d\Phi(\p) 
  {1\over \om_1\om_2\om_3} 
  \left[ \om_1\delta(\om - \om_1 - \om_2 - \om_3)\right. \nn
&&\hspace{4cm} +\left. \om\delta(\om + \om_1 - \om_2 -\om_3) \right]\, .
\ea
As discussed in \cite{ourwork}, this expression agrees with the imaginary
part of the quantum self-energy to order $\lambda^2$ for $\omega \ll T$, where
the Bose-Einstein distribution function becomes
\be
n(\om) \simeq {T\over \om} \gg 1\, .
\ee
From Eq.~(\ref{gamc}) one easily obtains the plasmon damping rate,
\be
\gamma = {1\over 2m} \Gamma^{(2)}({\bf 0},m) 
= {1\over 1536\pi} {\lambda^2 T^2\over m} \left(1 -
  {12\over \pi^2}{m\over \L} + \cO{\left(m\over \L\right)^2}\right)\;.
\ee
Except for the corrections $\cO{m\over \L}$ the result agrees with the
quantum result to leading order in $\lambda$ \cite{AaSm}. The dependence
on the infrared cutoff $\L$ controls the size of corrections to the
infrared limit.

To assess the size of the quantum corrections one has to compute the
contribution of Fig.~\ref{fig:selfen}d to the damping rate (diagram c. does 
not contribute to the imaginary part, diagram e. containes less IR dominant 
propagators and therefore gives a smaller contribution). The correction to the
classical self-energy $\bar\Pi$ reads (cf. appendix)
\begin{equation}
  \delta\bar\Pi(y_1,y_2)=\frac{\lambda^2}2\,\bar
  G_{T,\Lambda}^2(y_1-y_2)\left[ G^2_{++}(y_1-y_2)-G^2_{+-}(y_1-y_2) \right].
\end{equation}
After some algebra one finds for the correction to the imaginary part of the 
on-shell self-energy (${\bf p}=0,\,\omega=m$),
\ba
\delta\bar\Gamma^{(2)}({\bf 0},m)&=&\frac{\lambda^2}2\pi T\int d\Phi({\bf 0})\ 
\frac1{\omega_1}\delta(m+\omega_1-\omega_2-\omega_3) \nn
&=& \frac{\lambda^2T^2}{768\pi}\cdot\frac6{\pi^2} \left(\beta m \ln
  \frac m{2\Lambda} + \cO{\beta\L}\right)\; .
\ea
At high temperatures $m^2\sim \lambda T^2$, i.e., the quantum corrections
are ${\cal O}(\sqrt{\lambda}\ln\lambda)$. This implies that the
classical limit gives a meaningful approximation to the quantum theory
only to the leading order in $\lambda$ \cite{ourwork}.

\section{Summary}

We have shown that in scalar quantum field theories at high temperature the 
behaviour of low frequency modes is described by an effective classical
theory. Corrections to this classical limit are controlled by the dependence
of observables on an infrared cutoff $\L$ which is introduced to separate 
soft modes from hard modes.

Integrating out the hard modes yields an effective theory which depends
on the infrared cutoff $\L$ and on a classical background field $\phi_\L(x)$,
the expectation value of the field operator in a coherent state containing
soft modes only. The thermal Green functions of the low energy effective
theory factorize in products of expectation values of the field operator
up to corrections of order $\hbar\beta\L$. This loss of quantum coherence
is due to the interactions with the soft modes in the thermal bath. The 
behaviour of soft modes is described by classical statistical field theory.

The separation of soft and hard modes has been carried out at some fixed
initial time $t_i$. Due to interactions with the thermal bath soft and
hard modes will mix at later times. The implications for the behaviour
of correlation functions at large times remain to be studied.

Our analysis also illustrates some of the difficulties which have to be
faced in calculations of real-time correlation functions in gauge theories   
at high temperature. The separation of soft and hard thermal modes is
complicated by gauge invariance and in the cases of interest the gauge
coupling is rather large, so that the magnetic mass scale $g^2 T$ is not well
separated from the temperature $T$. Hence, it is not clear to us how
accurately existing classical numerical simulations for real-time correlation 
functions in gauge theories approximate the quantum theory.\\   

We would like to thank D.~Boyanovsky, M.~L\"uscher, A.~Patk\'os and 
J.~Pol\'onyi for helpful discussions and comments. This work has been 
partially supported by the Hungarian NSF under contract OTKA-T22929.

\section*{Appendix}

The contour integral in Eq.~(\ref{lamnl}) is most easily evaluated in the
two-component formalism which is obtained by taking the limits $t_i 
\rightarrow - \infty$, $t_f \rightarrow \infty$. Then only the contributions 
from the parts $C_+$ and $C_-$ of the Keldysh-type contour in Fig.~1 remain.

The corresponding propagators read (cf.~\cite{LaWe}) 
\ba
G_{++}(x_1-x_2) &=& \Theta(t_1-t_2) G^>(x_1-x_2) + 
                      \Theta(t_2-t_1) G^<(x_1-x_2) \nn 
                &=& \left(G_{--}(x_1-x_2)\right)^*\; ,\\
G_{+-}(x_1-x_2) &=& G^<(x_1-x_2)\ =\ \left(G_{-+}(x_1-x_2)\right)^*\; ,
\ea
where
\ba
G^{>(<)}(x) &=& \int_\k\ G^{>(<)}(\k,t)\ e^{i\k\x}\; ,\\
G^>(\k,t) &=& e^{-i\om_\k t} + 2 {n(\om_\k)\over \om_\k}
              \cos{\om_\k t} \nn 
          &=& \left(G^<(\k,t)\right)^*\; .
\ea

The interactions of the two fields $\phi_+$ and $\phi_-$ which correspond to
the two branches $C_+$ and $C_-$, respectively, are given by
\ba
S_I &=& - \int dt \int d^3x {\lambda\over 4!} \left((\phi_+ + \phi_\L)^4 -
         (\phi_- + \phi_\L)^4\right) \nn
&=& - \int dt \int d^3x \lambda \left({1\over 6}(\phi_+ - \phi_-)\phi_\L^3 
+ {1\over 4}(\phi_+^2-\phi_-^2)\phi_\L^2\right. \nn
&&\hspace{2cm} \left. + {1\over 6}(\phi_+^3 - \phi_-^3)\phi_\L + 
{1\over 24}(\phi_+^4-\phi_-^4)\right)\; ,
\ea
where
$\phi_\L$ is the background field introduced in Sect.~2.

For the contribution of diagramm \ref{fig:kernels}$i$ one now obtains
\ba
&&{\lambda^2\over 4}\phi_\L(x)\ \int_c dx'\ G_\L(x,x')^2\ 
                        \hphi(x')_\L^2 \nn
&& ={\lambda^2\over 4}\phi_\L(x)\ \int d^4x'\ 
                 \left(G_{++}(x-x')^2-G_{+-}(x-x')^2\right)\phi_\L^2(x')\nn
&& ={\lambda^2\over 4}\phi_\L(x)\ \int dt' \Theta(t-t')
\int d^3x' \left(G^>(x-x')^2 - G^<(x-x')^2 \right)\phi_\L^2(x')\; .\nonumber
\ea
Inserting the expressions for the propagators $G^>$ and $G^<$ and performing
a partial integration with respect to $t'$ one obtains the result 
Eq.~(\ref{nlocal}).


\newpage
\begin{thebibliography}{99} 

\bibitem{ShapRub} 
For a discussion and references, see\\ 
V.~A.~Rubakov and M.~E.~Shaposhnikov, 
{\em Usp. Fiz. Nauk.} {\bf 166} (1996) 493

\bibitem{GriRub}
D.~Yu.~Grigoriev and V.~A.~Rubakov, {\em Nucl. Phys.} {\bf B229} (1988) 67

\bibitem{ChSisim} 
J.~Ambj{\o}rn and A.~Krasnitz, {\em Phys. Lett.} {\bf B362} (1995) 97;
preprint NBI-HE-97-18, hep-ph/9705380

\bibitem{ASY}
P.~Arnold, D.~Son and L.~Yaffe, {\em Phys. Rev.} {\bf D55} (1997) 6264 

\bibitem{paris}
G.~Parisi, {\em Statistical Field Theory} (Addison-Wesley, New York, 1988)

\bibitem{BoMcSm} 
D.~B\"odeker, L.~McLerran and A.~Smilga, {\em Phys. Rev.} {\bf D52} (1995) 4675

\bibitem{Bod} 
D.~B\"odeker, {\em Nucl. Phys.} {\bf B486} (1997) 500

\bibitem{heinz}
R.~R.~Parwani, {\em Phys. Rev.} {\bf D45} (1992) 4695; {\em ibid.} {\bf D48}
(1993) 5965 (E);\\
E.~Wang and U.~Heinz, {\em Phys. Rev.} {\bf D53} (1996) 899

\bibitem{AaSm} 
G.~Aarts and  J.~Smit, {\em Phys. Lett.} {\bf B393} (1997) 395;
preprint THU-97-15, hep-ph/9707342

\bibitem{ourwork} 
W.~Buchm\"uller and A.~Jakov\'ac, {\em Phys. Lett.} {\bf B407} (1997) 39

\bibitem{jak}
A.~Jakov\'ac, {\em Classical Limit in Scalar QFT at High Temperature}, in 
Proc. of {\em Strong and Electroweak Matter '97} (Eger, Hungary, May 1997),
to appear; DESY 97-148, hep-ph/9708229

\bibitem{weert}
B.~J.~Nauta and Ch.~G.~van Weert, hep-ph/9709401

\bibitem{Lifsic} 
L.~D.~Landau and E.~M.~Lifshitz, {\em Statistical Physics} 
(Pergamon Press, London, 1959)

\bibitem{wetter}
N.~Tetradis and C.~Wetterich, {\em Nucl. Phys.} {\bf B398} (1993) 659

\bibitem{pietro}
M.~D'Attanasio and M.~Pietroni, {\em Nucl. Phys.} {\bf B472} (1996) 711

\bibitem{LaWe} 
N.~P.~Landsman and Ch.~G.~van~Weert, {\em Phys. Rep.} {\bf 145} (1987) 141

\bibitem{ItzZub} 
C.~Itzykson and J.-B.~Zuber, {\em Quantum Field Theory} 
(McGraw-Hill, New York, 1980)

\bibitem{NiSe} 
A.~J.~Niemi and G.~W.~Semenoff, {\em Ann. of Phys.} {\bf 152} (1984) 105

\bibitem{zinn}
J.~Zinn-Justin, {\em Quantum Field Theory and Critical Phenomena}
(Clarendon Press, Oxford, 1993)

\bibitem{BoyVeg} 
D.~Boyanovsky, H.~J.~de Vega, R.~Holman, D.-S.~Lee
  and A.~Singh, {\em Phys. Rev.} {\bf D51} (1995) 4419

\bibitem{JakPat} 
A.~Jakov\'ac, {\em Phys. Rev.} {\bf D53} (1996) 4538;\\
A.~Jakov\'ac and A.~Patk\'os, {\em Nucl. Phys.} {\bf B494} (1997) 54


\end{thebibliography}
\end{document}